%% file: ms-revised.tex
\documentclass[manuscript]{aastex}

\def\feka{Fe K$\alpha$}
\def\chandra{{\it Chandra}}
\def\xmm{{\it XMM-Newton}}
\def\suzaku{{\it Suzaku}}
\def\swift{{\it Swift}}
\def\asca{{\it ASCA}}
\def\hst{{\it HST}}

\def\sax{{\it BeppoSAX}}

\def\rosat{{\it ROSAT}}

\def\lum{erg s$^{-1}$}
\def\flux{erg cm$^{-2}$ s$^{-1}$}
\def\nh{cm$^{-2}$}
\def\arcsec{$^{\prime\prime}$}
\def\deg{$^{\circ}$}

\def\ltsima{$\; \buildrel < \over \sim \;$}
\def\simlt{\lower.5ex\hbox{\ltsima}} 
\def\gtsima{$\; \buildrel > \over \sim \;$}
\def\simgt{\lower.5ex\hbox{\gtsima}} 

\def\3c{3C~445}

\input{psfig.sty}

\begin{document}

\title{The remarkable X-ray spectrum of the Broad-Line Radio Galaxy 3C~445}

\author{R. M. Sambruna} 
\affil{NASA/GSFC, Code 661, Greenbelt, MD 20771
(rms@milkyway.gsfc.nasa.gov)}

\author{J. N. Reeves}
\affil{NASA/GSFC, Code 662, Greenbelt, MD 20771 and 
Astrophysics Group, School of Physical \& Geographical Sciences, Keele
University, Keele, Staffordshire ST5 5BG, UK} 

\author{V. Braito}
\affil{NASA/GSFC, Code 662, Greenbelt, MD 20771 and 
Department of Physics \& Astronomy, Johns Hopkins University, Baltimore, MD 21218}

\begin{abstract}

We present the results of the analysis of an archival 15~ks \xmm\
observation of the nearby ($z$=0.057) radio-loud source \3c, optically
classified as a Broad-Line Radio Galaxy. While the RGS data are of
insufficient quality to allow a meaningful analysis, the EPIC data
show a remarkable X-ray spectrum. The 2--10~keV continuum is described
by a heavily absorbed (N$_H \sim 10^{22}-10^{23}$ \nh) power law with
photon index $\Gamma \sim 1.4$, and strong ($R \sim 2$) cold
reflection. A narrow, unresolved \feka\ emission line is detected,
confirming previous findings, with EW $\sim$ 120~eV. A soft excess is
present below 2~keV over the extrapolation of the hard X-ray power
law, which we model with a power law with the same photon index as the
hard power law, absorbed by a column density N$_H=6 \times 10^{20}$
\nh\ in excess to Galactic. A host of emission lines are
present below 2~keV, confirming previous indications from \asca, due
to H- and He-like O, Mg, and Si. We attribute the origin of the lines
to a photoionized gas, with properties very similar to radio-quiet
obscured AGN. Two different ionized media, or a single stratified
medium, are required to fit the soft X-ray data satisfactorily. The
similarity of the X-ray spectrum of \3c\ to Seyferts underscores that
the central engines of radio-loud and radio-quiet AGN similarly host
both cold and warm gas.

\end{abstract}

{\sl Subject Headings:}{Galaxies: active --- galaxies: radio -- 
galaxies: individual --- X-rays: galaxies}

\section{Introduction}

The X-ray emission from AGN is a powerful tool to investigate the
structure and physical conditions of the matter in the proximity of
the central supermassive black hole. In particular, sensitive X-ray
spectroscopy has been very successful in disentangling the
contributions from warm and cold matter in Seyferts. At soft X-rays,
more than 50\% of these sources exhibit complex intrinsic
absorption/emission features suggesting the presence of photoionized
gas (Crenshaw et al. 2003), containing a significant fraction of the
accretion mass. In a handful of sources, blueshifted absorption
features were observed, indicating an outflow from the nucleus with
quasi-relativistic velocities, $v/c$ \gtsima 0.1 (Chartas et al. 2002,
Pounds et al. 2003, Reeves, O'Brien, \& Ward 2003; see also Braito et
al. 2007 and references therein). The same absorbing gas is thought to
be responsible for the soft X-ray emission lines observed in type-2
sources (Guainazzi et al. 2005; Turner et al. 1997), via scattering or
reflection (Netzer 1996). Reflection features at hard X-rays (\feka\
line, Compton hump) indicate reprocessing by cold gas in an accretion
disk (George \& Fabian 1991, Nandra et al. 1997), and provide a way to
explore the innermost regions around the black hole where
gravitational effects are most important.  A prominent, broad \feka\
emission line at 6.4 keV, together with a hump peaking around 20--30
keV (the ``Compton hump'') indicate reprocessing of the optical-UV
emission in colder gas arranged in an accretion disk. A pc-scale
molecular torus encases the accretion flow, spurring orientation-based
unification models.

In contrast, the inner regions of radio-loud AGN are much more poorly
studied, because of the relative rarity and distance of these
sources. Broad-Line Radio Galaxies (BLRGs) are by now thought to
exhibit weaker reflection features and flatter X-ray continua than
Seyfert 1s (e.g., Sambruna, Eracleous, \& Mushotzky 2002, and
references therein). No evidence for warm absorbers was detected
so far in bright BLRGs at energies \ltsima 2~keV with \rosat\ and
\asca\ (Sambruna et al. 1999, S99 in the following; Reynolds 1997),
\sax\ (Grandi, Malaguti, \& Fiocchi 2006), and with \chandra\ and
\xmm\ (Gliozzi et al. 2007; Lewis et al. 2005; Ogle et al. 2005;
Ballantyne et al. 2004). A recent 120~ks \suzaku\ observation of
3C~120 shows a featureless continuum at soft X-rays, attributed
to the radio jet (Kataoka et al. 2007). It may thus seem that,
contrary to Seyferts, the central engines of BLRGs are devoid of
warm gas. 

However, there are reasons to expect the presence of a medium in BLRGs
and other radio-loud AGN. For example, centrifugally-driven winds,
lifting matter off the disk's surface and channeling it down the
magnetic field, are a proposed scenario for the origin of relativistic
jets (Blandford \& Payne 1982); at favorable orientations, these winds
would lead to observable discrete absorption/emission features at soft
X-rays (K\"onigl \& Kartje 1994).  A dense environment can also be
responsible for slowing down the inner jet via mass entrainment
inferred in more beamed sources (Georganopoulos \& Kazanas
2003). Unification models for radio-loud sources also postulate the
presence of a warm, scattering gas to explain type-2 sources (Urry \&
Padovani 1995).

The study of the gas environs of BLRGs would benefit from X-ray
observations of a source seen at optimal angles, such as to allow us
to peek into the central engine while at the same time minimizing the
jet contribution. This opportunity is offered by \3c, a nearby
(z=0.057) and X-ray bright, F$_{2-10~keV} \sim 7
\times 10^{-12}$ \flux, radio-loud source optically classified as a
BLRG but with a large inclination angle as inferred from the radio
(\S~2). Previous \asca\ observations suggest a heavily absorbed X-ray
continuum (Sambruna et al. 1998), similar to obscured radio-quiet
AGN. 

Here we present and discuss an archival 15~ks \xmm\ observation of
\3c, which confirms its type-2 X-ray spectrum and shows
the presence of a host of soft X-ray emission lines, for the first
time in a radio-loud AGN. The paper is structured as follows. The
sources properties and the observations are summarized in
\S\S~2 and 3. The results of the spectral fits are presented in \S~4,
while discussion and conclusions follow in \S\S~5 and 6. Throughout
this paper, a concordance cosmology with H$_0=71$ km s$^{-1}$
Mpc$^{-1}$, $\Omega_{\Lambda}$=0.73, and $\Omega_m$=0.27 (Spergel et
al. 2003) is adopted.

\section{3C~445}

The nearby (z=0.057) radio galaxy \3c\ has an FRII radio morphology
(Kronberg, Wielebinski, \& Graham 1986), with a linear extension up to
10\arcmin. With a steep radio spectrum between 2.7 and 4.8 GHz
($\alpha_r$=0.7) and a core-to-lobe intensity ratio R=0.039 (Morganti,
Killeen, \& Tadhunter 1993), the source is clearly
lobe-dominated. From the projected radio size of the source Eracleous
\& Halpern (1998) infer an inclination $i>$ 60\deg. Using the ratio of
the radio fluxes of the approaching (South) and receding (North) jets
($\sim$ 7.7; Leahy et al. 1997), we derive an upper limit to
the inclination angle $i<$ 71\deg\ (for p=3 in eq. A10 of Urry \&
Padovani 1995). In the following we will assume $i$=60\deg. 

Broad emission lines were observed in the optical unpolarized flux
(Eracleous \& Halpern 1994; Corbett al. 1998), leading to its
classification as a BLRG. The spectrum is very red, and steepens
dramatically at UV energies (Crenshaw, Peteron, \& Wagner 1988). It
has been suggested that the large IR emission is the result of the
reprocessing of the optical and UV emission by circumnuclear dust
(Elvis et al. 1984). The optical continuum is polarized (Brindle et
al. 1990). The amount of reddening derived from the large Balmer
decrement (H$\alpha$/H$\beta \sim$ 8; Crenshaw et al. 1988,
Osterbrock, Koski, \& Phillips 1976) and by the large
Pa$\alpha$/H$\beta$ ratio (5.6; Rudy \& Tokunaga 1982) is E(B-V)=1
mag. For a standard dust-to-gas conversion ratio a column density N$_H
\sim 5 \times 10^{21}$ \nh\ is derived. This is one order of magnitude
larger than the Galactic column density in the direction to the
source, N$^{Gal}_H=5.33 \times 10^{20}$ \nh, derived from 21~cm
measurements (Murphy et al. 1996). Based on these properties, and on
the limits to the viewing angle, we suggest we are seeing 3C~445
almost edge-on. Thus, the beamed jet is not expected to contribute
significantly to the emission from the source. In fact, for
$i$=60\deg\ and a bulk Lorentz factor $\Gamma_J \sim 10$, the Doppler
factor of the jet is $\delta \sim 0.2$. Thus, the jet emission is
actually de-boosted ($F_{J,obs}=\delta^{3+\alpha}F_{J,intr}$, where
$\alpha>0$ is the radio energy index; Urry \& Padovani 1995).

At odds with its classification as a type-1 source, \3c\ exhibits an
X-ray spectrum very similar to Seyfert 2s. Previous \rosat\ and \asca\
observations (Sambruna et al. 1998, S98 in the following), while of
insufficient quality to allow a full spectral decomposition, indicated
a heavily absorbed (N$_H \sim 10^{23}$ \nh) continuum above 3~keV,
plus a narrow \feka\ line with EW $\sim$ 250 eV, and a soft excess
modeled with an unabsorbed power law. From a later reanalysis of the
data, S99 noted weak emission lines below 2~keV in the
non-simultaneous \rosat\ and \asca\ data, attributed to ionized
elements from O to Si. The limited quality of the data, however,
prevented a detailed analysis of the lines and their origin.

Moreover, \3c\ was detected with the Burst Alert Telescope onboard
\swift\ in the energy range 15--150~keV (Tueller 2007, priv. comm.)
and with the PDS onboard \sax\ (Grandi, Malaguti, \& Fiocchi 2006;
Dadina 2007). However, \3c\ lies close (30\arcmin) to the cluster of
galaxies A2440 (z=0.094), and contamination of the hard X-ray flux by
the cluster can not be ruled out in the \sax\ data.  Nevertheless, the
broad-band \sax\ spectrum was fitted with a partial-covering,
dual-absorber model, yielding $\Gamma=1.7$ and N$_H
\sim 10^{23}$ \nh. These data also gave the first measure of the
reflection continuum (after subtraction of the expected cluster
contribution); the latter is weakly constrained, with reflection
fraction $R\sim 3$ and large uncertainties. Here, $R=\Omega/2\pi$ is
the fraction of the reflector's solid angle seen by the illuminating
continuum (George \& Fabian 1991).

Thus, previous X-ray observations of \3c\ provided tantalizing clues
that the 0.5--10~keV emission from this BLRG is complex. Moreover,
analysis of the \xmm\ field of view of \3c\ reveals the presence of a
nearby (1.3\arcmin) soft X-ray AGN (Grandi et al. 2004), which was not
resolved at the \asca\ and \sax\ poor angular resolutions. Thus, the
\xmm\ observation presented here yields the very first high-quality
X-ray spectrum of this source.

\section{Observations}

\xmm\ observed \3c\ on December 6, 2001 for a total exposure of
23.8~ks. After screening the data the net exposure, which includes
correction for deadtimes, was 15.3~ks with the EPIC pn, 21.3~ks with
MOS1, and 21.2~ks with MOS2. The count rate of the source in 0.4--10
keV is 0.59 $\pm$ 0.006 c/s with the pn and 0.18 $\pm$ 0.003 c/s with
both MOS CCDs. With the RGS we collected a total of 280 counts with both
instruments in the energy range 0.4--2~keV. 

The pn, MOS1, and MOS2 cameras were operating in Small Window mode,
with the Thin filter applied. The \xmm\ data have been processed and
cleaned using the latest Science Analysis Software (SAS ver 7.0) and
analyzed using standard software packages (FTOOLS ver.  6.1, XSPEC
ver. 11.3).  In order to define the threshold to filter for
high-background time intervals we extracted the 10--12~keV light curve
and filtered out the data when the light curve is 2$\sigma$ above its
mean. The inspection of this light curve shows that there was no
flaring activity during this observation.  For our analysis only
events corresponding to pattern 0--12 for the EPIC MOS and pattern
0--4 for the pn were used. EPIC pn source spectra were extracted using
a circular region of 30\arcsec, while background data were extracted
from a circular region with radius 1\arcmin\ centered at $\sim$
2.5\arcmin\ from the source. This is sufficient to include most of the
PSF, while excluding the 1.3\arcmin\ X-ray source 1WGA J2223.7-0206
(Grandi et al. 2004). The EPIC MOS1 and MOS2 data were extracted using
a source extraction region of 30\arcsec\ radius and two background
regions with identical size (30\arcsec), selected on the nearby CCDs.
Response matrices and ancillary response files at the source position
have been created using the sas tasks
\verb+arfgen+ and \verb+rmfgen+.

The EPIC MOS1 and MOS2 data were summed for the spectral fits and the
background-subtracted spectrum binned to have at least 20 counts in
each energy bin; the background-subtracted EPIC pn data were binned to
have at least 30 counts per bin, to validate the use of the $\chi^2$
statistics.  The spectra were fitted within \verb+XSPEC+ v. 11.3.2ad.
Finally, we checked that there were no discrepancies between spectra
extracted with this selection and spectra extracted with pattern 0.

\section{Spectral Fitting Results}

We first fitted the EPIC pn and MOS spectra with a simple power law
plus Galactic absorption. As expected this was not a good fit
($\chi^2_r=3.71/556$), yielding a photon index
$\Gamma=-0.26$. Figure~\ref{plaw} shows the data in the source's
rest-frame, with the model removed for clarity. At energies above
3~keV, a curved continuum is visible indicating heavy absorption. The
\feka\ line at 6.4~keV is also apparent. At soft X-rays, several emission
features can be seen in the energy range 0.4--2~keV, attributed to
lighter elements from O to Si. In the remainder of this section, all
energies are in the source's rest-frame unless otherwise specified.

\subsection{The Continuum}

\begin{figure}[h]
\centerline{
\includegraphics[scale=.6,angle=-90]{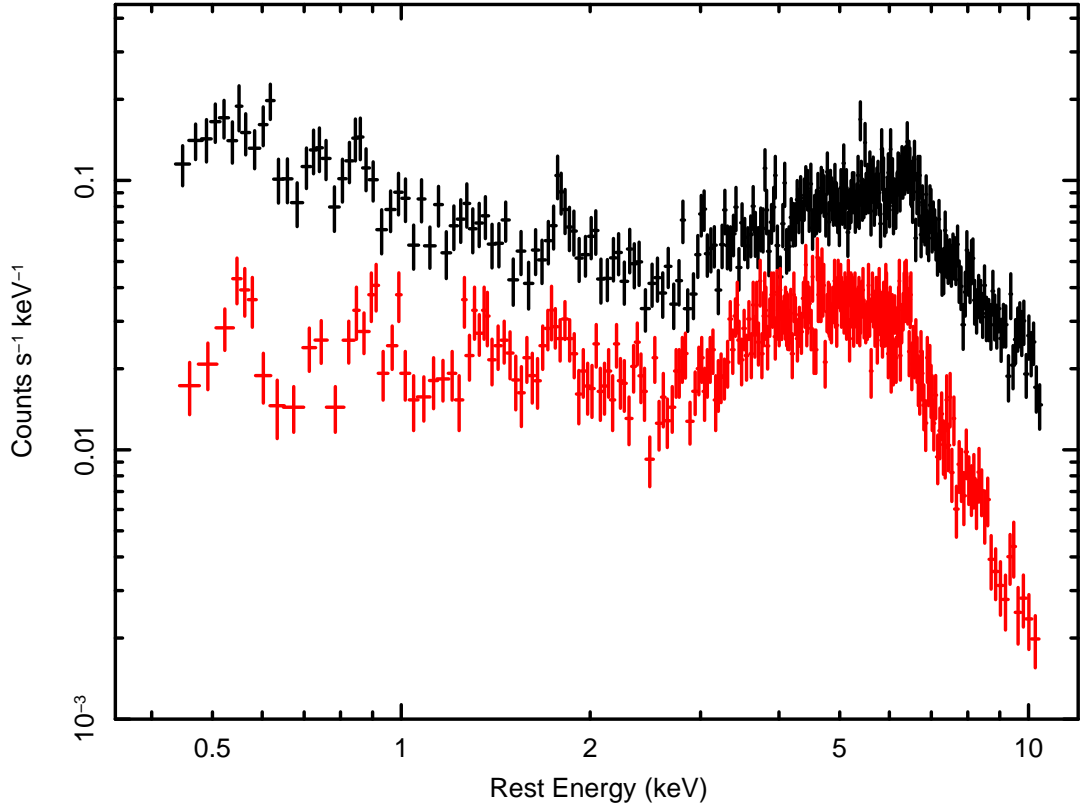}}
\caption{\footnotesize
The EPIC spectra of \3c, from an archival 15~ks observation. The
spectrum with higher flux is the pn, while the remaining spectrum is
the sum of MOS1 and MOS2. The data were fitted with the baseline model
of a single power law with Galactic absorption; for clarity the model
is not shown here. A complex X-ray spectrum is apparent, with a bumped
continuum above 3~keV, the \feka\ line at 6.4~keV, and several
emission lines in 0.4--2~keV. }
\label{plaw}
\end{figure}

We first concentrated on modeling the continuum. At the harder
energies, \gtsima 3~keV, the bumpy shape of the continuum
(Fig.~\ref{plaw}) suggests heavy absorption, as in the previous \asca\
observations (S98). Fitting the EPIC data in 3--10~keV with a single
power law plus free column density indeed yields significant
absorption, N$_H^1=1.3 \pm 0.1 \times 10^{23}$ \nh, with 
$\Gamma=1.27 \pm 0.13$ and $\chi^2_r$=1.2/386. The residuals of
this model clearly show an emission line feature at 6~keV
(Fig.~\ref{plaw}), identified with the \feka\ line detected with
\asca\ (S98). Ignoring the energy range 5--7~keV, where possible
contributions from a broad relativistic \feka\ line are often seen in
Seyferts, yields an acceptable fit ($\chi^2_r=1.0/228$) but still an
unusually hard continuum, $\Gamma_h=1.02 \pm 0.15$, and excess column
density N$_H=(9.6 \pm 1.4) \times 10^{22}$ \nh. Thus, the flat
continuum slope above 3~keV is not an artifact of a complex profile of
the \feka\ line.

For the remaining of the spectral fits, we now consider the full
0.4--10~keV band with the 5--7~keV bins added back. Addition of the
lower energy bins shows that the 0.4--3~keV flux lies 2--3 orders of
magnitude above the extrapolation of the absorbed power law,
indicating the presence of a soft continuum component. This confirms
the previous findings from \asca\ and \rosat\ joint fits, which
required the presence of a steeper power law at the softer energies
possibly related to scattering of the nuclear radiation by ambient
gas/dust (S98). We thus added a second power law component to the fit,
absorbed by a column density N$_H^2$ and with photon index $\Gamma_s$
tied to the index of the hard power law, $\Gamma_s=\Gamma_h$. This fit
yields $\chi^2_r=1.57/553$, still poor primarily because of
unaccounted emission lines\footnote[1]{If narrow Gaussians are added
to model the X-ray lines (Table~2), the $\chi^2_r=1.1/540$.}. The
photon index is now $\Gamma_s=\Gamma_h=1.22 \pm 0.07$, with an
absorption column density N$_H^2 \sim 6 \times 10^{20}$ \nh\ above the
Galactic value acting mainly on the soft power law. Leaving the soft
and hard indices free to vary yields $\Gamma_s=1.44 \pm$ 0.11 and
$\Gamma_h=0.90 \pm$ 0.13, both still quite flat, and
$\chi^2_r=1.54/552$. 

The presence of the \feka\ emission line (see below) and the \sax\ PDS
detection suggest that addition of a reflection continuum might be
necessary. We thus added cold reflection modeled with \verb+pexrav+
(Magdziarz \& Zdziarski 1995), where the abundances were fixed to
solar values and the inclination angle $i$ to 60\deg. The free
parameters of the fit were the reflection fraction, $R$, and the
normalization. The fit is greatly improved, $\Delta\chi^2$=117, and
yields very strong reflection, $R \sim 3$. While not physical, taken
at face value this would indicate a strong reflection component
dominating the emission at energies \gtsima 8~keV.

However, if we add a third layer of absorption to the model,
specifically, a third power law with photon index tied to $\Gamma_h$
and column density N$_H^3$, the strength of the reflection component
decreases, $R \sim 2$, and N$_H^3 \sim 4 \times 10^{22}$ \nh.  Due to
the limited EPIC bandpass and the complexity of the spectrum there is
no simple way based on the data to decide whether the 2--10~keV
continuum is dominated by the reflection component, or if instead
multiple layers of absorption are present. Here we adopt the latter
model, specifically, three distinct cold absorbers and reflection
(Table~1); as we discuss later, there is no evidence for this source
to be Compton-thick and thus dominated by reflection in the 2--10~keV
energy band. Our approved \suzaku\ observations will allow us to
discriminate among the various possibilities, which clearly have
different physical implications, and measure the intrinsic AGN flux.

The best spectral decomposition of the 0.4--10~keV continuum is
obtained with a model including: three powerlaws, all with the same
photon index $\Gamma \sim 1.4$, and two layers of cold absorption,
N$_H=10^{22}-10^{23}$ \nh; and strong ($R \sim 2$) cold
reflection. Figure~\ref{bestfit} shows the data fitted with the above
model (and with the inclusion of the emission lines, see below), and
the best-fit model itself. The best-fit parameters and their 90\%
uncertainty are reported in Table~1, where the observed fluxes and
intrinsic (absorption-corrected) luminosities are also given. The
covering fractions of the three cold absorbers are 84\% (N$_H^1$), 4\%
(N$_H^2$), and 12\% (N$_H^3$).

To exclude that the line-like residuals at soft X-rays are due to 
inappropriate continuum modeling, we attempted fitting the EPIC
spectrum with the baseline continuum model but using an ionized
reflector (\verb+reflion+) instead of the cold one. This model is
known for producing relatively strong soft X-ray emission lines for
intermediate ionization states (Ross, Fabian, \& Young 1999). The fit
with this model is worse by $\Delta\chi^2$=30. Basically, the model
fails to account simultaneously for the various ionization properties
at soft and hard X-rays. Specifically, accounting for the Mg and Si
lines would require a strong high-ionization Fe line at 6.7--6.9~keV,
which is not observed.

Note that the measured continuum photon index is rather hard, $\Gamma
\sim 1.4$, flatter than usually observed in Seyferts or other
BLRGs. From a \suzaku\ observation of 3C~120, Kataoka et al. (2007)
derive $\Gamma \sim 1.7$, similar to what was obtained from \asca\
observations of BLRGs (S99). The limited EPIC bandpass, together with
the presence of complex absorption, most likely conjures to produce an
apparently hard continuum. Fixing $\Gamma$ to 1.7, in fact, yields an
equivalent fit, with an increase of the column densities of 20\%. This
reflects the well-known spectral degeneracy between slope and
absorption column over a limited bandpass ($<$ 10~keV). 

The total observed fluxes, F$_{0.4-2~keV} \sim 2 \times 10^{-13}$ and
F$_{2-10~keV} \sim 7 \times 10^{-12}$ \flux, are clearly lower limits
to the true AGN fluxes in these energy bands, and fully consistent
with the fluxes measured with \asca\ and \rosat\ (S98). A 2--10~keV
observed flux lower by a factor 2 was measured in earlier
\sax\ observations\footnote[2]{Note, however, that the \sax\ and
\asca\ datasets contained a (presumably small) contribution from the
1.3\arcmin\ source; thus, the factor 2 variations between the
\sax\ and \xmm\ epochs is clearly a lower limit.} (Grandi et
al. 2006). Thus, there is evidence that the medium-hard X-ray flux
varied significantly over the timescale of almost 2 years between the
\sax\ and \xmm\ observing epochs. 

Extrapolating the best-fit model at harder X-ray energies, we find a
15--100~keV observed flux of F$_{15-100~keV} \sim 5 \times 10^{-11}$
\flux. This is a factor 2 larger than the BAT flux threshold, and
consistent with the claimed detection.

\subsection{The Fe line region} 

In Figure~\ref{feline}a we plot the residuals of the baseline best-fit
continuum model discussed above (a power law with three absorbers plus
cold reflection), zooming in the \feka\ emission line region. These
residuals were obtained by fitting the full-band EPIC data minus the
5--7~keV region, and then adding back the 5--7~keV for the plot. 

The prominent emission line at 6.4~keV is the \feka\ emission line,
present both the pn and MOS data (Fig.~\ref{plaw}). Adding a Gaussian
improves the fit significantly, $\Delta\chi^2$=57 for 3 additional
parameters. The line is narrow, $\sigma=70^{+49}_{-69}$~eV, and
unresolved at $>$ 95\% confidence; its observed Equivalent Width
against the total continuum is EW $\sim$ 120~eV (Table~2), while the
EW with respect to the reflection continuum only is
$524^{+173}_{-165}$~eV. Adding a broad Gaussian to the narrow line
does not improve the fit and its width is completely
unconstrained. Figure~\ref{feline}b shows the residuals after fitting
a narrow Gaussian.

The measured EW with respect to the total observed continuum is by
itself suggestive of a strong reflection component (George \& Fabian
1991). Indeed, the intensity of the \feka\ line depends at first order
on the amount of reflection, the iron abundance, and the inclination
angle. Taking into account the limited bandpass of \xmm, the well
known degeneracy between R and $\Gamma$, and the complex absorption in
this source, the amount of reflection cannot be directly derived from
the continuum itself. However, assuming an inclination angle of
60\deg, solar abundances, and an intrinsic $\Gamma$ of 1.4 the ratio
between the \feka\ emission line and the reflected component is
expected to be $1.4\times 10^{-2}$.  From our best fit we derive a
ratio of about $\sim0.01$, close to the expected value.  On the other
hand, if the normalization of the Fe line and the reflection continuum
are linked together the residuals in 5--7~keV are flat and no
additional broad or narrow components to the \feka\ emission line are
required.

In both Figure~\ref{feline}a-b, an absorption feature is present
around 6.8~keV. The dip was modeled with an inverted Gaussian, leading
to a modest ($\Delta\chi^2$=10) improvement of the fit, significant at
$\sim$ 97\% confidence from the F-test. The fitted energy and EW are
E=6.87 $\pm$ 0.09~keV and EW=40$^{+22}_{-25}$~eV. If the absorption
feature is real, the closest candidate for its identification would be
the 1$\rightarrow$2 transition of FeXXV at 6.701~keV, blueshifted by a
modest amount, $v\sim 0.02c$. The EW would imply a column density
N$_H\sim 10^{22}$\nh. However, the feature significance is
hindered by the choice of the underlying continuum model; a stronger
reflection component would reduce the significance of the dip. This is
because of the correlation between $\Gamma$ and $R$: a stronger
reflection component would allow the slope to become slightly steeper,
decreasing the contrast between the continuum and the absorption
features. 

\subsection{Soft X-ray Emission Lines}

At softer energies, several emission lines are present in 0.4--1~keV
and further around 2~keV (Figure~\ref{plaw}). We interpret these
features as emission lines due to the lighter elements from O to
Si. Table~2 lists the observed line energies, fluxes, EWs, and
their identifications. 

The lack of a strong FeL complex at 0.9--1~keV provides evidence
against an origin of the X-ray lines in collisionally ionized gas,
such as the galaxy ISM. Adding a thermal model (\verb+apec+) to the
best-fit model yields $\chi^2_r$=1.09/544, and leaves line-like
residuals at 0.7 and 0.8~keV, and at 1.3 and 1.8~keV. The fitted
temperature is $kT \sim 0.12$~keV, with abundances $\sim 0.7$
solar. Previous \rosat\ PSPC data showed the source is consistent with
being point-like (S98), although higher-resolution \chandra\ data are
needed to confirm this result.

A strong possibility is an ionized medium, most likely photoionized by
the intense nuclear light, as commonly observed in Seyfert
galaxies. Indeed, we detect two emission features at 0.74 and 0.87~keV
which could be interpreted as the Radiative Recombination Continuum
(RRC) of OVII and OVIII, respectively (Table~2). The RRC features link
unequivocally the origin of the emission lines to a plasma
photoionized by the AGN, generally at a temperature of a few eV
(Liedahl \& Paerels 1996). At the EPIC resolution the RRC features are
unresolved. In the case of OVII RRC, a contribution to the flux from
the nearby ($\Delta\lambda=0.299$ \AA) FeXVII 3s-2p line can not be
excluded; usually, however, the OVII RRC is much brighter (factor 2 or
more) than FeXVII (Kinkhabwala et al. 2002).

We thus added a photoionization model component to describe the soft
X-rays, using the code \verb+XSTAR+ (Bautista \& Kallman 2001). The
latter describes the emission lines expected from a medium whose
physical conditions are summarized in the ionization parameter
$\xi=L_{ph}/nr^2$ ergs cm/s, where $L_{ph}$ is the luminosity of the
photoionizing continuum, $n$ the gas density, and $r$ its distance from
the nucleus. Other parameters are the element abundances relative to
solar, the column density of the gas, N$^W_H$, and the
normalization. Since the model is clearly underconstrained by the
data, we fixed the elemental abundances to their best-fit values
(consistent with solar within the uncertainties), and the column
density to N$^W_H=10^{21}$ \nh, similar to the columns observed in
Seyfert 1s (Reynolds 1997). Leaving N$^W_H$ free to vary does not
improve or change the fit, and furthermore, the value is
unconstrained. Only the ionization parameter and the normalization
were left free to vary.

We find that the EPIC data require two separate media, one with
ionization parameter $\log \xi_1 \sim$ 1.9, accounting for most of the
O and Ne lines, and a second one with $\log \xi_2 \sim$ --0.09,
responsible for the Mg and Si lines. The addition of a second ionized
medium is significant at $>$ 99.9\% confidence ($\Delta\chi2=22$ for 2
additional parameters). As discussed elsewhere (NGC~4507; Matt et
al. 2004) the need for two ionization parameters does not imply that
the media are physically separate; in fact, they could represent two
different zones of the same cloud. The parameters of the best-fit
model with \verb+XSTAR+ are reported in Table~1, and the model is
shown in Figure~\ref{bestfit}. The individual lines fluxes and EWs are
listed in Table~2.

Albeit the low quality of the RGS data (\S~3), we inspected the
RGS1, 2 spectra to search for the brightest emission lines. The data
are very noisy and affected by large (50\% or more) errorbars; only a
slight hint (1$\sigma$) for a possible line at 0.57~keV is present,
with a total of 10 counts in the feature. Thus, we will not discuss
the RGS data any further.  

In conclusion, an archival 15~ks EPIC observation of \3c\ provides
evidence for a very complex X-ray spectrum. In particular, several
X-ray emission lines are present in 0.4--2~keV, most likely due to a
photoionized gas. To our knowledge, this is the first time that soft
X-ray emission lines are detected in a radio-loud AGN. The presence of
X-ray lines from \3c\ was previously claimed based on non-simultaneous
\rosat\ and \asca\ spectra (S99), which also indicated a heavily
absorbed continuum. In addition to confirming these features, the EPIC
data also provide evidence for a strong reflection continuum and/or
several layers of cold absorption of the power law continuum.

\section{Discussion}

Overall, the X-ray spectrum of \3c\ is remarkably similar to a Seyfert
2, at odds with its classification as type-1 AGN. The EPIC spectrum
indicates a strong, albeit uncertain without hard X-ray observations,
reflection component which is expected to dominate the continuum above
10~keV, and indeed \3c\ was detected with the BAT and PDS
experiments. Given the large inclination angle of the radio source
(\S~2), it is quite possible that the Compton hump is due to
reflection off a medium along the line of sight (the torus wall?). The
\feka\ line is narrow, with no evidence for a broad component linked
to the disk. Its EW $\sim$ 120~eV is consistent with being produced by
both transmission and reflection through a cold medium with column
N$_H$
\gtsima $10^{23}$ \nh\ (Turner et al. 1997). 

Alternatively, the requirement in the EPIC data of an unusually strong
reflection component is alleviated by adding multiple cold absorbers
to the continuum model (\S~4.1). This scenario would imply the
presence of different layers of absorption, which could possibly
totally obscure the primary nuclear X-ray continuum in the EPIC
bandpass. To this regard, we tested the possibility that \3c\ is
Compton-thick, i.e., there is yet another, unaccounted absorber, N$_H$
\gtsima $10^{24}$ \nh, using the diagnostic diagram of Bassani et
al. (1999). This is a plot of the \feka\ EW versus the thickness
parameter T, defined as the ratio of the total 2--10~keV flux of the
source and the intrinsic [OIII] flux (Bassani et al. 1999). In this
plot, Compton-thick sources tend to occupy the region of large EW and
small T values. 

The observed [OIII] flux of \3c\ is F$_{obs}[OIII]=1.7 \times
10^{-13}$ \flux\ (Tadhunter et al. 1998). Correcting for the reddening
using the observed Balmer decrement (\S~2), the intrinsic flux is
F$_{obs}[OIII]=3 \times 10^{-12}$ \flux. From the 2--10~keV flux in
Table~1, the T parameter is T=2.2. This value and the \feka\ EW $\sim$
120~eV (Table~1) place \3c\ in the region of AGN with absorption
column densities N$_H \sim 10^{22}-10^{23}$ \nh\ in Figure~1 of
Bassani et al. (1999), near the Seyfert 1s locus. For comparison, the
two radio-loud sources in the sample of Bassani et al., Cyg~A and
NGC~6251, have EW=380~eV and T=22, and EW=390~eV and T=2,
respectively; they are located among (NGC~6251) or close (Cyg~A) to
the Seyferts. Thus, there is no evidence based on the presently
available data that \3c\ is Compton-thick. 

Perhaps the most striking result of the EPIC data analysis is the
detection of X-ray emission lines below 2~keV. Again, this is similar
to what generally found in obscured radio-quiet AGN (Turner et
al. 1997, Guainazzi et al. 2005), where photoionization is generally
dominant but with a non-negligible contribution from resonant
scattering (Kinkhabwala et al. 2002). Indeed, the EPIC spectrum of
\3c\ is very similar to the Compton-thin Seyfert 2 NGC~4507, where
emission lines from ionized O, Ne, Mg, and Si (as well as a strongly
absorbed continuum) were detected with \xmm\ with similar EWs (Matt et
al. 2004).  As in the case of NGC~4507, two ionized emitters or a
single emitter with a range of ionization parameters are required for
\3c. The ionization parameters we measure for \3c\ (Table~1) are
within the range ($\xi^1$) or slightly lower ($\xi^2$) than typically
found for type-2 Seyferts (e.g., Kinkhabwala et al. 2002).

We conclude that the central engine of \3c\ hosts a warm gas with
properties similar to radio-quiet obscured AGN, which is responsible
for the observed soft X-ray lines, most likely through scattering. The
location of this ``warm mirror'' remains unknown with the present
data. Using \chandra\ and \hst\ [OIII] images, Bianchi, Guainazzi, \&
Chiaberge (2006) showed that the soft X-ray emission of obscured
radio-quiet AGN is extended and spatially coincident with the
NLR. Unfortunately, no \chandra\ images are yet available to test this
hypothesis in the case of \3c. Based on the fact that we can see the
soft X-ray lines, we can at least locate the warm mirror further out
from the cold gas affecting the continuum emission (Matt et al. 2004). 
Another possibility is an outflow extending vertically above the plane
of the torus, particularly intriguing given the radio-loud nature of
\3c\ and the presence of a radio jet (see below). 

From the EPIC best-fit model, we calculated an upper limit to the
distance $r$ of the warm gas from the black hole, using the gas
parameters from Table~1. Under the assumption that the gas forms a
thin shell, $\Delta r/r <1$, for a nuclear ionizing (corrected for
absorption) luminosity L$_{ph}=L_{0.4-100~keV} \sim 3 \times 10^{44}$
\lum. Assuming that the medium has a density similar to the BLRs
($n=10^9$ cm$^{-3}$), $r=0.02-0.2$ pc. At these distances the gas
velocity should be $v \sim$ 5,000--16,000 km/s for a black hole mass
$10^9$ M$_{\odot}$. If instead $n=10^3$ cm$^{-3}$, consistent with the
NLRs, then $r>10$ pc and $v < 500$ km/s.  These scenarios can be
tested with high-resolution X-ray spectroscopy which will resolve the
individual lines yielding the density and location of the photoionized
gas, and will measure intrinsic velocity dispersions/energy shifts.

The covering fraction of the soft X-ray emitting gas can also be
calculated from the normalization (luminosity) of the photoionized
emission modeled by XSTAR. The photoionized matter may exist in a
shell of gas around the photoionizing AGN source, covering a fraction
($f$) of $4\pi$\,sr$^{-1}$ solid angle. From the XSTAR code, the
normalization ($k$) of a component of photoionized gas is related to
the covering fraction by $f=D_{\rm kpc}^{2} k / L_{ph}$, where $D_{\rm
kpc}$ is the distance to \3c\ in kpc and $L_{ph}$ is the ionizing
luminosity in units of $10^{38}$ \lum. From above,
$L_{ph}=3\times10^{6}$ \lum, while
$D^{2}=6\times10^{10}$\,kpc$^{2}$. From the XSTAR fits, the luminosity
of the photoionized emission corresponds to $L_{photo}=6\times10^{41}$
\lum, with approximately equal contribution from the high and low
ionization zones, while the XSTAR normalization corresponds to
$k=2\times10^{-5}$ ph cm$^{-2}$ s$^{-1}$ (with an uncertainty of
$\sim50$\%) for an assumed column density N$_H^W=10^{21}$ \nh. Thus
this translates into a covering fraction $f \sim$ 40\% for
N$_H^W=10^{21}$ \nh. If the assumed column density is higher,
N$_H^W=10^{22}$ \nh, then the required covering fraction is
correspondingly lower, e.g., $f
\sim4$\%. Given that by definition $f<1$, the lower limit on the
column density of the soft X-ray emitting gas is $>4\times10^{20}$
\nh\ at a maximum distance of $<5$\,kpc.

In any event, a more accurate estimate of the covering fraction of the
photoionized gas cannot be made without a measurement of the ionized
column of gas in absorption against the direct continuum
emission. This may be achievable in the forthcoming long Suzaku
observation of 3C 445, through refinement of the measurement of the
possible Fe\,XXV absorption line near 6.8\,keV, or from the possible
detection of softer absorption lines from He/H-like Si and S near 2
keV.

The detection of soft X-ray lines in \3c\ has implications for other
BLRGs in the context of unification models. In radio-quiet Seyferts,
the ``warm mirror'', is commonly unified with the ionized outflows
producing the absorption features observed in the soft X-ray spectra
of type-1 objects, the ``warm absorber'' (Netzer 1996). Assuming that
the general unification scenario of Seyferts holds for radio-loud AGN
as well (Urry \& Padovani 1995), and that \3c\ is fairly
representative of other radio-loud AGN, one can infer that a warm
absorbing gas should be present in the more aligned counterparts of
\3c, producing absorption features at soft X-rays. If this is indeed
the case, one may wonder why these features were not detected in
BLRGs.

The answer is twofold. On one hand, there are only four bright BLRGs
traditionally known: 3C~120, 3C~382, 3C~390.3, and 3C~111, all of them
observed extensively at X-rays. Both superluminal sources 3C~120 and
3C~111 lie at low Galactic latitudes, preventing a detailed study of
the emission below 2~keV. In addition, 3C~120 displays a variable soft
excess attributed to the jet (Kataoka et al. 2007); unless the
absorber subtends a large solid angle to the jet (e.g., entrained
matter), any absorption feature will be weak and difficult to detect
against the beamed non-thermal continuum. In 3C~382 the soft excess is
also variable (Barr \& Giommi 1992), and contributions to the soft
X-rays from the diffuse circumnuclear thermal emission (Gliozzi et
al. 2007) could mask weak absorption features.

On the other hand, deep X-ray observations of BLRGs with a
signal-to-noise ratio comparable to Seyfert 1s were never performed in
the past. High-resolution grating observations of BLRGs so far exist
only for 3C~382 (HETGS, 120~ks) and 3C~390.3 (RGS, 90~ks). In the
former, two emission lines at $\sim$ 0.89 and 1.04~keV were marginally
detected (Gliozzi et al. 2007) but their origin is not clear due to
the limited quality of the MEG data.

The best candidate BLRG for the detection of absorption features is
arguably 3C~390.3. In fact, this galaxy has a low (N$_H=3.7 \times
10^{20}$ \nh) Galactic column, a relatively large X-ray flux
(F$_{2-10~keV} \sim 10^{-11}$ \flux), and a powerlaw continuum
(S99). Previous X-ray observations of 3C~390.3, however, provide
controversial evidence for the presence of absorption. A variable
column density of cold gas was observed in multiepoch X-ray spectra of
this BLRG (Grandi et al. 1999), while a 40~ks \asca\ spectrum provided
evidence for an absorption edge at 0.74~eV with optical depth $\tau
\sim 0.3$, identified eith the OVII edge at 0.739~eV (S99). However,
no apparent absorption but instead weak emission lines were detected
at soft X-rays in a 90~ks \xmm\ EPIC and RGS exposure (Lewis et
al. 2007, in prep.), confirmed in our recently acquired \suzaku\ data
(Sambruna et al. 2007, in prep.).

We now comment on the apparent discrepancy of the X-ray spectrum with
the optical type-1 classification of \3c. This difficulty may be
circumvented in the ``clumpy hydromagnetic wind'' torus model of
Elitzur \& Shloshman (2007; see also K\"onigl \& Kartje 1994, Everett
2005). In this model, the torus is not a continuous, donut-like
structure but is composed of clouds distributed around the equatorial
plane of the AGN (see Fig.~3 in Elitzur 2006); the torus extends in
the inner regions beyond the dust sublimation radius, where the clouds
become atomic and ionized yielding broad optical lines and warm X-ray
absorption. For a given number of clouds and their distribution on the
equatorial plane, classification of an AGN as type 1 or 2 depends on
whether the line of sight intercepts enough obscuring clouds.

The multiwavelength properties of \3c\ can be reconciled within this
scenario. The inclination angle of \3c\ from the radio is larger
(\S~2) than the current estimates of the average opening cone of the
``torus'', $\sim 30-45$\deg\ (Schmitt et al. 2001), implying we might
be looking at the nucleus of \3c\ through significant number of
clouds. Our line of sight could be such to be obscured by matter in
the outer, colder (N$_H \sim 10^{23}$ \nh) molecular clouds of the
toroidal distribution, while at the same time intercepting the broad
emission lines from innermost ionized BLR clouds if the latter are
lifted at some height above the plane (Elitzur \& Shlosman 2007). The
BLR clouds could also be responsible for the lower column density,
N$_H^1 \sim 6 \times 10^{20}$ \nh, and, via scattering, the soft X-ray
continuum (Table~1). The external dustier clouds would be responsible
for the observed continuum and H$\alpha$ polarization, as well as for
the strong IR continuum emission.

The clumpy wind scenario provides a viable scenario also for the X-ray
emission lines from \3c. K\"onigl \& Kartje (1994) suggested that the
X-ray warm absorber/emitter observed in Seyferts coincides with clouds
uplifted from the surface of the accretion disk by the magnetic field
in a hydromagnetically driven sub-relativistic outflow (Blandford \&
Payne 1982). Indeed, X-ray observations at moderate-to-high spectral
resolution of Seyferts 1/2 detected signatures of winds. A large
proportion of these sources exhibit soft X-ray absorption lines
blueshifted relative to systemic, indicative of low-velocity, $\sim$
1,000 km/s, outflows; at larger luminosities, fast outflows ($v$
\gtsima 0.1c) were observed in a small number of systems on the basis
of FeK absorption features (see Braito et al. 2007 and references
therein).

In the clumpy wind model the conditions for developing such an outflow
become particularly favorable toward the inner parts of the disk
(Elitzur \& Shlosman 2007). From this perspective, it is tantalizing
to speculate that the X-ray emission lines observed from
\3c\ originate from an inner outflow component, perhaps related to the
formation of the radio jet itself (Blandford \& Payne 1982; K\"onigl
\& Kartje 1994; Proga et al. 2000). The EPIC data, however, do not
provide conclusive evidence for the presence of an outflow in \3c. A
weak absorption feature is observed at 6.8~keV, but its significance
is hindered by the choice of the underlying continuum model (\S~4.2).

\section{Conclusions}

Using an archival 15~ks \xmm\ spectrum we have shown that the X-ray
emission from the nearby BLRG \3c\ is quite complex. In particular,
its properties are remarkably similar to those of type-2 radio-quiet
AGN, with a heavily absorbed continuum and a strong reflection
component, a narrow \feka\ emission line, and several soft X-ray lines
consistent with reflection/scattering off a warm ``mirror''.

The soft X-ray lines are of particular interest, as this is the first
time that such features are detected in a BLRG and indeed in a
radio-loud AGN. Again, their properties are similar to radio-quiet
Seyfert 2s. The most likely origin of the lines is from a photoionized
gas close to the nucleus. Follow-up X-ray spectroscopy of \3c\ with
higher resolution is strongly encouraged to determine the location and
density of the warm medium.

If unification models hold, and if \3c\ is representative of
radio-loud AGN, one would expect to observe absorption features from
the same ``warm mirror'' in BLRGs more favorably oriented close to the
line of sight. So far, such features have eluded detection.  Future
high-quality X-ray observations of radio-loud AGN of both type 1 and 2
are needed to investigate the gas content of these systems.

\acknowledgements

This research has made use of data obtained from the High Energy
Astrophysics Science Archive Research Center (HEASARC), provided by
NASA's Goddard Space Flight Center, and of the NASA/IPAC Extragalactic
Database (NED) which is operated by the Jet Propulsion Laboratory,
California Institute of Technology, under contract with the National
Aeronautics and Space Administration. We are grateful to the referee
for helpful comments.


\clearpage
\input{tab1}

\clearpage
\input{tab2}

\clearpage


\newpage


\begin{figure}[h]
\includegraphics[scale=.5,angle=-90]{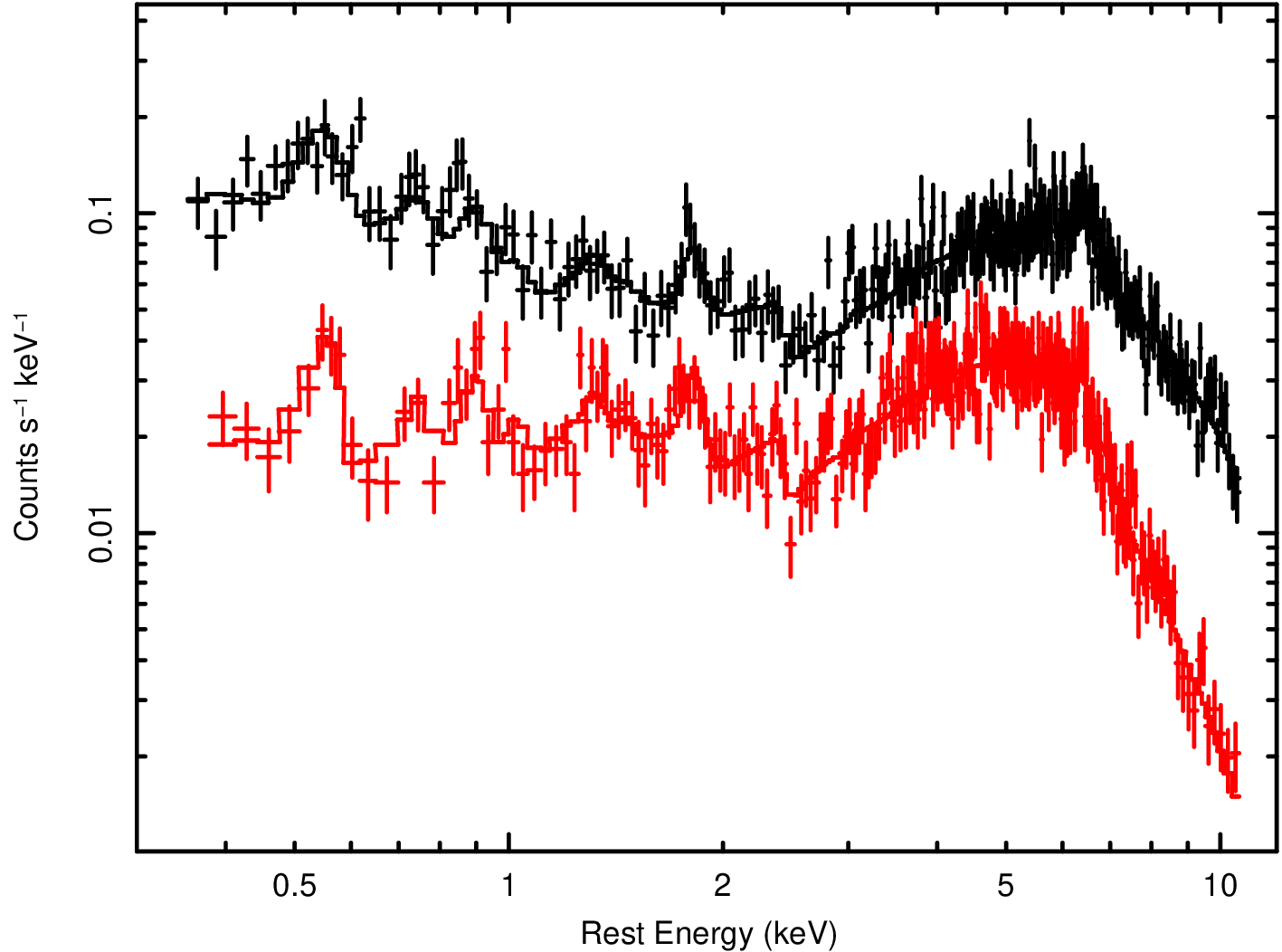}
\includegraphics[scale=.5,angle=-90]{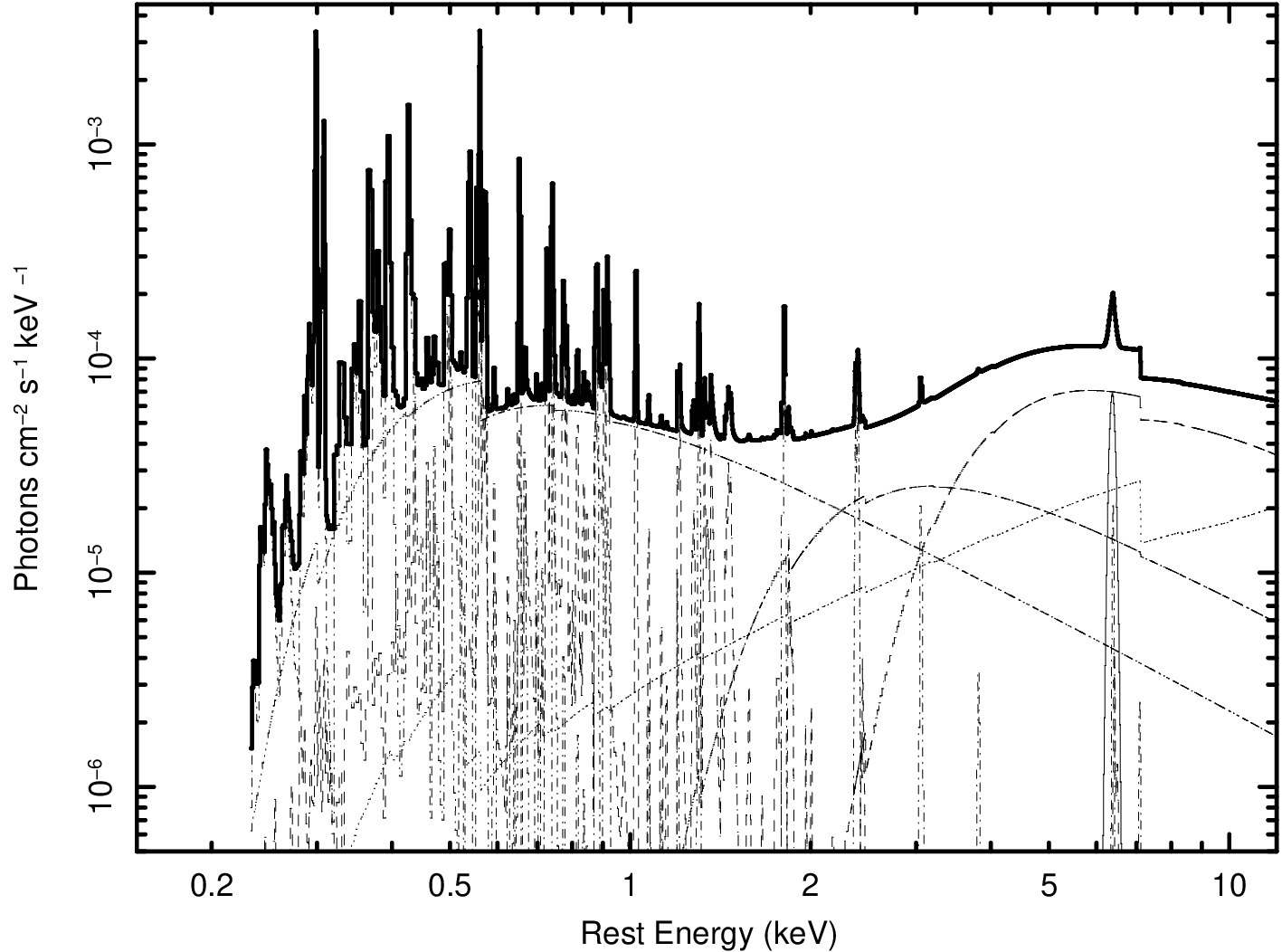}
\caption{\footnotesize EPIC pn and MOS1+2 data (top panel) and
best-fit model (bottom panel). See Table 1 and 2, and text.  
}
\label{bestfit}
\end{figure}

\newpage


\begin{figure}[h]
\includegraphics[scale=.5,angle=-90]{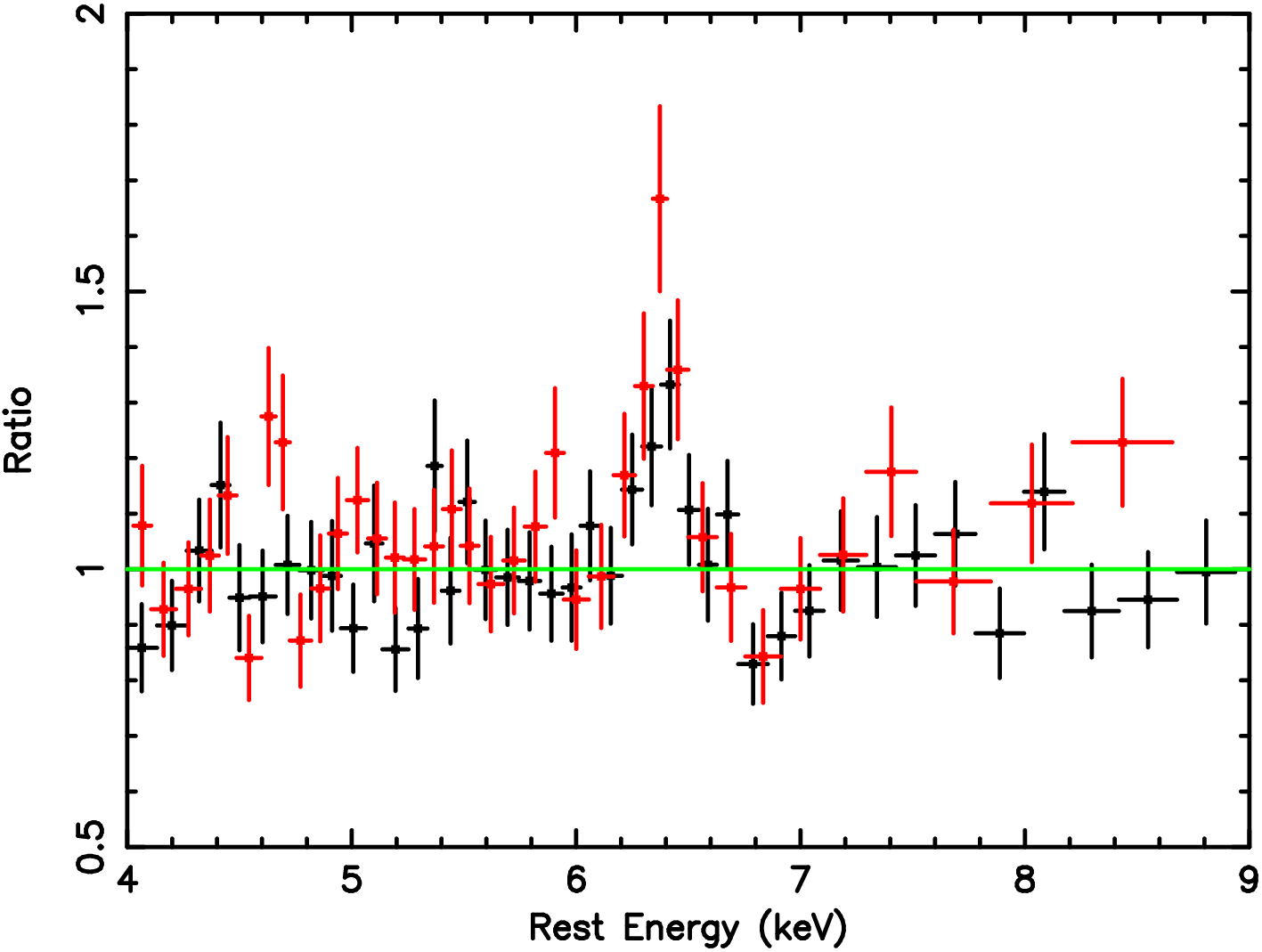}
\includegraphics[scale=.5,angle=-90]{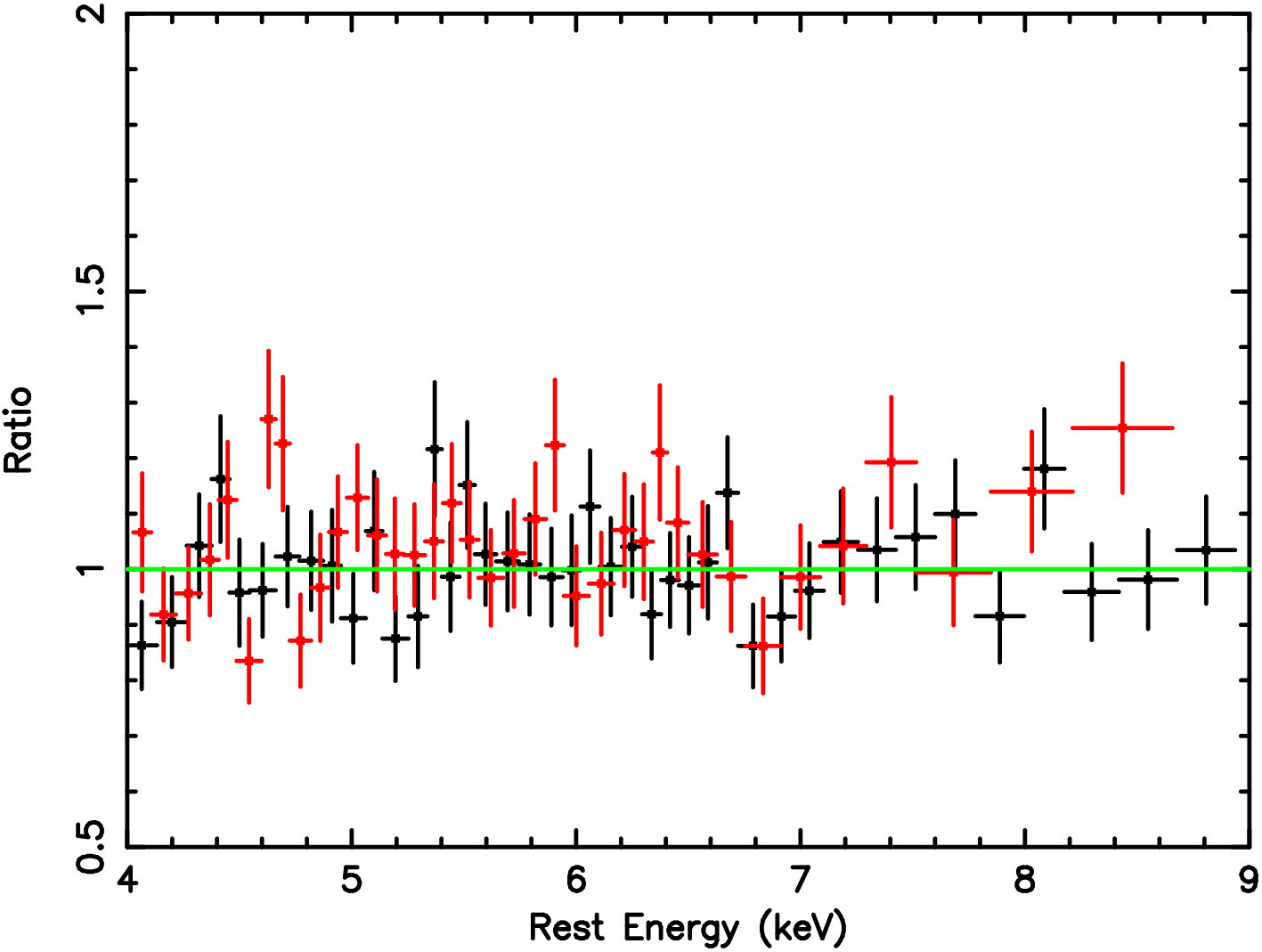}
\caption{\footnotesize {\it (Top, a):} Residuals of a fit to the EPIC
data of \3c\ with the best-fit continuum model (Table~1) in the region
of the \feka\ line. A narrow line is visible, together with an
absorption dip at 6.8~keV. {\it (Bottom, b):} Residuals of the same
model, but with a narrow Gaussian line added at 6.4~keV.  } 
\label{feline} 
\end{figure}

\newpage

\end{document}

%% file: tab1.tex


\scriptsize
\begin{center}
\begin{tabular}{lcc}
\multicolumn{3}{l}{{\bf Table 1: EPIC Best-fit Parameters$^{\dagger}$}} \\
\multicolumn{3}{l}{   } \\ \hline
& &  \\
Parameter & & \\ 
& & \\ \hline
N$_H^1$  && 22.5$^{+6.2}_{-4.3} \times 10^{22}$ \nh \\
N$_H^2$  && 5.8$^{+10}_{-5.7} \times 10^{20}$ \nh  \\
N$_H^3$  && 4.2$^{+2.4}_{-1.7} \times 10^{22}$ \nh \\
$\Gamma$ && 1.39 $\pm 0.16$ \\
&& \\
R$_{refl}$   && 1.9$^{+3}_{-1.8}$ \\
E$_{fold}$   && 100~keV \\
cos$i$       && 0.5 fix \\
&& \\
E$_L$      && 6.38 $\pm$ 0.03 keV \\
$\sigma_L$ && 70$^{+49}_{-69}$ eV \\
EW$^{\star}$     && $120^{+30}_{-40}$ eV \\
&& \\
$\log\xi^1$ && 1.86$^{+0.74}_{-0.68}$ erg cm/s\\ 
$\log\xi^2$ && --0.09$^{+0.18}_{-0.11}$ erg cm/s \\ 
N$_H^W$     && $1 \times 10^{21}$ \nh\ fix \\
&& \\
$\chi^2_r$  && 1.02/541 \\
&& \\
F$_{0.4-2~keV}$ && $2 \times 10^{-13}$ \flux \\
F$_{2-10~keV}$ && $6.7 \times 10^{-12}$ \flux \\
&& \\
L$_{0.4-2~keV}$ && $1.1 \times 10^{44}$ \lum \\
L$_{2-10~keV}$ && $8.6 \times 10^{43}$ \lum \\
&&\\ \hline

\end{tabular}
\end{center}

\noindent {\bf Notes:} $\dagger$=The best-fit model consists of: three
power laws with tied photon index absorbed by N$_H^1$, N$_H^2$, and
N$_H^3$, plus cold reflection, plus two warm emitters modeled with the
code \verb+XSTAR+. All components are absorbed by Galactic N$_H^G=5.33 
\times 10^{20}$ \nh; $^{\star}$=The EW is calculated with respect to the total 
observed continuum. The fluxes are observed, and the luminosities
intrinsic (absorption-corrected).

\normalsize

%% file: tab2.tex


\scriptsize
\begin{center}
\begin{tabular}{lccc}
\multicolumn{3}{l}{{\bf Table 2: X-ray emission lines}} \\
\multicolumn{3}{l}{   } \\ \hline
& &  \\
Energy (keV) & Flux ($10^{-6}$ ph cm$^{-2}$ s$^{-1}$)  & EW$\dagger$ (eV) & Identification \\ 
& & & \\ \hline
0.57 $\pm$ 0.01 & 31.0 $\pm$ 0.59 & 170 $\pm 30$  & OVII K$\alpha$ \\
&&&\\
0.73 $\pm$ 0.01 & 6.3 $\pm$ 2.6  & 49$^{+22}_{-19}$  &  OVII RRC \\
&&&\\
0.87 $\pm$ 0.01 & 6.8 $\pm$ 2.0 & 67 $\pm$ 20 &  OVIII RRC \\
&&&\\
1.34 $\pm$ 0.03 & 2.3 $\pm$ 1.2  &42 $\pm$ 20 &  Mg XI \\
&&&\\
1.78 $\pm$ 0.03 & 2.9 $\pm$ 1.2 & 73$ \pm 30$ &  Si XIII K$\alpha$ \\
& & &\\
6.38 $\pm$ 0.03 &  14.5 $\pm$ 0.4 & $120^{+30}_{-40}$ & \feka \\
&&& \\ \hline

\end{tabular}
\end{center}

\noindent {\bf Notes:} $\dagger$=The EWs are measured against the
total observed continuum at their respective energies.

\normalsize